\documentclass[twocolumn]{aastex62}
\usepackage{amsmath}

\usepackage{xcolor}
\usepackage{multirow}

\shorttitle{A Database of Rocky Exoplanet Albedo Spectra}
\shortauthors{Smith et al.}

\begin{document}

\title{Utilizing a Database of Simulated Geometric Albedo Spectra for Photometric Characterization of Rocky Exoplanet Atmospheres}
\correspondingauthor{Adam Smith}
\email{adjsmith@nmsu.edu}

\author[0000-0002-9156-9651]{Adam J. R. W. Smith}
\affiliation{Southeastern Universities Research Association, 1201 New York Ave., NW, Suite 430, Washington, DC 20005}
\affiliation{Center for Research and Exploration in Space Science and Technology, NASA Goddard Space Flight Center, Greenbelt, MD 20771}
\affiliation{NASA Goddard Space Flight Center 8800 Greenbelt Road Greenbelt, MD 20771, USA}

\author[0000-0002-8119-3355]{Avi M. Mandell}
\affiliation{NASA Goddard Space Flight Center 8800 Greenbelt Road Greenbelt, MD 20771, USA}

\author[0000-0002-2662-5776]{Geronimo L. Villanueva}
\affiliation{NASA Goddard Space Flight Center 8800 Greenbelt Road Greenbelt, MD 20771, USA}

\author[0000-0001-7912-6519]{Michael Dane Moore}
\affiliation{Business Integra Inc., 6550 Rock Spring Dr., Suite 600, Bethesda, MD 20817}
\affiliation{NASA Goddard Space Flight Center 8800 Greenbelt Road Greenbelt, MD 20771, USA}

\received{May 1, 2020}
\revised{August 8, 2020}
\accepted{September 2, 2020}

\begin{abstract}
    In anticipation of future flagship missions focused on the goal of achieving direct imaging of rocky exoplanets, we have developed a database of models to help the community examine the potential spectral characteristics of a broad range of rocky planet atmospheres. Using the publicly available Planetary Spectrum Generator (PSG), we have computed a grid of 141,600 rocky exoplanet geometric albedo spectra across a 7-dimensional parameter space. Using this grid, we have performed a color-color analysis seeking to identify the most useful near-ultraviolet and red or near-infrared photometric follow-up channels to combine with a green-optical (discovery) spectral channel. We found that a combination of filters at 0.4 $\mu$m, 0.58 $\mu$m, and $\sim 0.8 \mu$m were able to distinguish between atmospheres with moderate-to-high concentrations of four different dominant absorbing constituents, given at least 10 hours of observation on a star at 10 parsec with a 15-meter-class space telescope; however, more moderate abundances similar to those of Solar System rocky bodies would be more challenging to detect. We recommend that future missions seeking to characterize directly imaged rocky exoplanets by colors alone further consider multi-band photometry as a first discriminator for planetary characteristics.
\end{abstract}


\section{Introduction}

Since the first discovery of a planet orbiting a main sequence star \citep{mayor_queloz_1995}, the study of exoplanets has grown tremendously. We have now begun to observe these planetary bodies in reflected light \citep[e.g.][]{chauvin2004giant}, and future large space telescopes currently under consideration are expected to make the first high-contrast reflected-light observations of rocky Earth-mass objects. In this paper, we use the NASA Planetary Spectrum Generator \citep[PSG, \url{https://psg.gsfc.nasa.gov};][]{villanueva2018planetary} to produce a grid of reflected-light spectra of rocky Earth-mass planets in Earth-like orbits around Sun-like stars with a range of simplified atmospheric abundance and aerosol profiles, with the goal of exploring and understanding the morphology of such spectra. Understanding the range of possible spectral morphologies will help future instrument designers and mission planners to efficiently plan exoplanet surveys with an eye towards characterization.

The field of exoplanetology has advanced rapidly in the past 25 years, from teasing out radial velocity detections \citep{mayor_queloz_1995} to atmospheric characterization \citep{charbonneau2002detection} and the success of sensitive transit detection surveys \citep{borucki2010kepler,ricker2014transiting,gillon2017seven}. Now, instruments such as the Very Large Telescope's Spectro-Polarimetric High-Contrast Exoplanet Research \citep[VLT-SPHERE;][]{beuzit2019sphere}, the Gemini South Telescope's Gemini Planet Imager \citep[GPI;][]{macintosh2015discovery}, the Palomar Observatory Project 1640 \citep{hinkley2011new} and the Subaru Telescope's Subaru Coronagraphic Extreme Adaptive Optics \citep[SCExAO;][]{jovanovic2015subaru} have successfully achieved direct imaging of dozens of brown dwarfs and other massive substellar objects. Future telescopes such as the Nancy Grace Roman Space Telescope \citep[RST, formerly WFIRST;][]{spergel2013wide} aim to enlarge this inventory. In addition, several large space telescope mission studies, such as the Large Ultra Violet / Optical / InfraRed Surveyor \citep[LUVOIR;][]{luvoir2019luvoir} and the Habitable Exoplanet Observatory \citep[HabEx;][]{gaudi2020habitable}, include as major mission objectives the direct imaging of Earth-sized exoplanets in Earth-like orbits around Sun-like stars in reflected (visible) light. The reflected-light spectrum of an exoplanet is a sensitive measure of the planet's atmospheric composition through photometric and spectroscopic measurements. Therefore, in advance of observations by RST and subsequent next-generation exoplanet imaging missions, recent studies \citep[such as][]{lupu2016developing,nayak2017atmospheric,batalha2018color,feng2018characterizing,smith2020detecting} have used a variety of sophisticated models to explore their potential scientific yield and develop tools to help characterize observed planets.

The use of state-of-the-art modeling techniques to produce large databases of model observations for comparison against real observations is common both within the exoplanet literature and across many fields of astronomy. For example, \citet{goyal2018library} published a model databases exploring giant planets in transmission spectroscopy, while \citet{batalha2018color} and \citet{macdonald2018exploring} computed publicly available grids of giant planet reflection spectra. \citet{allard2012models} and \citet{allard1996model} presented grids of atmospheres for low-mass stars, brown dwarfs and young giant planets. Many other databases of simulated observations exist, some going much further afield.

Future space missions will likely incorporate spatially-resolved spectroscopy (i.e. an integral field spectrograph, or IFS) to gather spectra of every source in the image plane sampled by a high-contrast exoplanet imaging instrument, and our model grid is produced at moderate resolution to accommodate future investigations of the science yield from spectroscopic analysis.  However, direct-imaging surveys of exoplanets will likely continue to utilize broadband photometric analysis methods even in the era of IFSs.  Photometric searches for Earth-like planets will most likely require an initial search without spectroscopy in order to detect background sources and efficiently characterize the system. These observations will either be photon-limited (assuming noiseless detectors) or detector-noise-limited (assuming more standard detectors), both of which benefit from photometric binning. The combination of multiple broadband photometric data points into ``colors'' has been the workhorse of astronomical classification from the early days of stellar physics up until today. Recently, some thought has been given towards applying similar methods to the characterization and study of rocky exoplanets like those in our own Solar System \citep{krissansen2016pale}, and there have been efforts to understand the efficiency of such a survey for giant planets with RST \citep{batalha2018color}.

In this study, we have applied these methods of database computation and color-color analysis towards the understanding of the science products of future exoplanet observation missions which aim to make direct observations of small, rocky exoplanets in reflected light. Acknowledging that exoplanets may vary wildly in composition, and that these variations can have dramatic consequences for observations, we first sought to parameterize and simplify the description of a rocky exoplanet atmosphere in order to reduce the dimensionality of our model grid. We then derived a physically meaningful range for each dimension, and explored the resulting parameter space with the NASA Planetary Spectrum Generator (PSG). Here, we present the resulting 141,600-model database, as well as some observations of the morphology of spectra contained therein and a simulated color-color analysis of the database.

In Section~\ref{sec:methods}, we discuss the tools and methods used in this project, including PSG. We also discuss the design of the parameter space for the grid. In Section~\ref{sec:database}, we present and validate our model database. In Section~\ref{sec:observation} we perform some basic photometric analysis using our database as a test case. Finally, in Section~\ref{sec:discussion}, we review and discuss the database and the findings from our photometric analysis, and suggest some future studies that could be performed by using this spectral database and avenues for its expansion.

\section{Methods}
\label{sec:methods}

In this section, we will discuss the techniques and choices employed in the development and calculation of our database of rocky exoplanet albedo models (hereafter ``the database''). In Subsection~\ref{subsec:design} we will describe the choices of parameters, and the chosen ranges of those parameters, that describe the parameter space explored by the database. In Subsection~\ref{subsec:PSG}, detail will be given of our choice of atmospheric forward model and its implementation in the computation of the database.

\subsection{Grid Design}
\label{subsec:design}

\begin{deluxetable*}{llll}[!]
    \tablecaption{Model Parameterization}
    \tablewidth{0pt}
    \tablehead{
    \colhead{Parameter Symbol} & \colhead{Description} & \colhead{Values Computed} & \colhead{Category}
    }
    \startdata
    C$_{abs}$ & Dominant absorbing chemical & [ H$_2$O, CH$_4$, CO$_2$, SO$_2$, O$_2$ ] & Atmosphere \\
    log(M$_{abs}$) & Dominant absorber mixing ratio & [ -7.0, -6.0, -5.0, -4.0, -3.0, -2.0, -1.0, -0.01 ] & Atmosphere \\
    H$_2$/(H$_2$ + N$_2$) & Background Gas Hydrogen-Nitrogen Ratio & [ 0, 0.02, 0.05, 0.15, 0.5 ] & Atmosphere \\
    log(P$_0$) [bar] & Surface Atmospheric Pressure & [ -1.0, -0.5, 0.0, 0.5, 1.0, 1.5, 2.0 ] & Atmosphere \\
    A$_s$ & Surface Albedo & [ 0.05, 0.2, 0.35, 0.5 ] & Bulk \\
    log($l$) [kg/m$^2$] & Column Cloud Mass & [ -3.0, -2.5, -2.0, -1.5, -1.0 ] & Cloud \\
    log(P$_t$) [bar] & Cloud top pressure & [ -2.0, -1.5, -1.0, -0.5, 0.0, 0.5, 1.0] & Cloud \\
    \enddata
    \tablecomments{The parameters, and ranges of values, used to generate the primary database of 136,000 model object geometric albedo spectra presented in this paper. The full parameter space described by these parameters is explored, except for models where P$_t$ $\geq$ P$_0$. In addition, an auxiliary grid of 5,600 geometric albedo spectra were generated for model atmospheres without clouds, for a total of 141,600 models.}
    \label{table:params}
\end{deluxetable*}

The problem of describing a planet's atmosphere and predicting its observable signatures can be quite complex, and to make it tractable at this stage we have sought to find and limit ourselves to the most important planetary parameters and determine a realistic range of values for each of these parameters. For a rocky planet, we might naturally divide these parameters into two groups: atmospheric parameters, describing the gaseous envelope surrounding the planet, and bulk parameters, describing the solid rocky surface and interior of the planet. Another division is useful when describing the atmosphere, as any aerosols, clouds, dust and haze suspended in the atmosphere suggest many specialized ``cloud'' parameters. As a first simplification, we chose to model our planets in one dimension, as additional dimensions increase computation time by orders of magnitude. Further, as is common in the literature \citep[i.e.][]{feng2018characterizing,smith2020detecting, ross2019simulated,lingam2019photosynthesis}, we reduced the surface characteristics to a simple isotropic grey reflecting surface\footnote{Recent publications have studied the repercussions of this choice. See Section~\ref{sec:discussion} for more discussion on this point.}, with a given numerical albedo value A$_s$ such that 0 $\leq$ A$_s$ $\leq$ 1.

It is also convenient to express mass in terms of $g$, or surface gravity, as this term (a function of mass and surface radius) is used in calculations of the atmospheric scale height. Direct-imaging measurements are much less sensitive to scale heights than transmission spectroscopy; as such, we chose to fix this gravity parameter at Earth-standard 9.8 m/s$^2$ in order to eliminate the parameter and allow us to explore other, more impactful parameters. Further, in our  model, the planet’s radius and orbital distance become only scalar coefficients on the overall planet’s insolation; therefore, we also fix the planet's radius at 1 R$_\Earth$. Surface temperature is set equal to the atmospheric temperature at the surface (described below), while internal temperature is deemed irrelevant for reasons explained above. Thus, we have reduced the bulk parameters to one, A$_s$.

We next consider atmospheric parameters. A full and complete description of a planetary atmosphere would require a host of parameters. Atmospheric temperature varies with altitude in all known planets, as do the mixing ratios of the atmosphere’s many constituent chemical species. Describing this variation in a tractable grid would require many parameters, yet using fixed vertical profiles typically captures the main molecular morphologies dominating a spectrum. Therefore we have constructed our atmosphere to be isothermal and vertically well-mixed, such that temperature and mixing ratios do not vary with altitude. In addition, we do not consider variations in the planet’s surface or atmospheric temperature, fixing both at 300 K. Tests which varied this parameter from 200K to 500K indicate it has no appreciable effect on continuum albedo, and only a slight effect on absorption features.

This leaves a simplified atmosphere described by its chemical mixing ratios and pressure profile. With our fixed planetary surface gravity, the pressure profile can be described completely with a single parameter: the surface atmospheric pressure, P$_0$. The atmospheric pressure at any altitude can then be computed from the scale height, $H = kT/mg$. Lastly, we chose to simplify the composition of our atmosphere from potentially hundreds of constituent chemical species to only three: two of which, H$_2$ and N$_2$, describe an inert background gas; and a third which represents the dominant absorbing molecule in the atmosphere. The composition of the background gas can be described by the ratio of H$_2$ to N$_2$, which we parameterize into the fraction H$_2$/(H$_2$+N$_2$) which lies between 0 and 1. The remainder of the atmosphere requires two parameters: the choice of dominant absorbing molecule, and the mixing ratio M$_{abs}$ of that molecule in the atmosphere. Thus we have described our atmosphere with four parameters: the non-continuous choice of dominant absorber C$_{abs}$, plus continuous variables of M$_{abs}$, H$_2$/(H$_2$+N$_2$), and P$_0$.

Finally, we consider the addition of aerosols. Like the atmosphere, the addition of aerosols opens us to many free parameters. These include the aerosols’ composition, altitude, vertical extent, and particle size and density. There may be multiple layers of clouds and/or hazes at different altitudes, and they may even have different compositions. In order to maintain computational feasibility, we were forced to discard many of these possibilities. We chose a single aerosol type - a simple representative ``white'' cloud - and assigned it an effective particle radius of 10 $\mu$m. These two choices nearly eliminate any variation in the aerosol’s properties as a function of wavelength. While this assumption breaks down at longer wavelengths, white clouds provide a reasonable approximation to both water and sulfuric acid clouds at near-UV, optical, and most near-IR wavelengths. In addition, with these choices, the cloud's optical thickness, $\tau$, becomes approximately proportional to the cloud deck’s density. 

\begin{figure}
    \centering
    \includegraphics[width=3.3in]{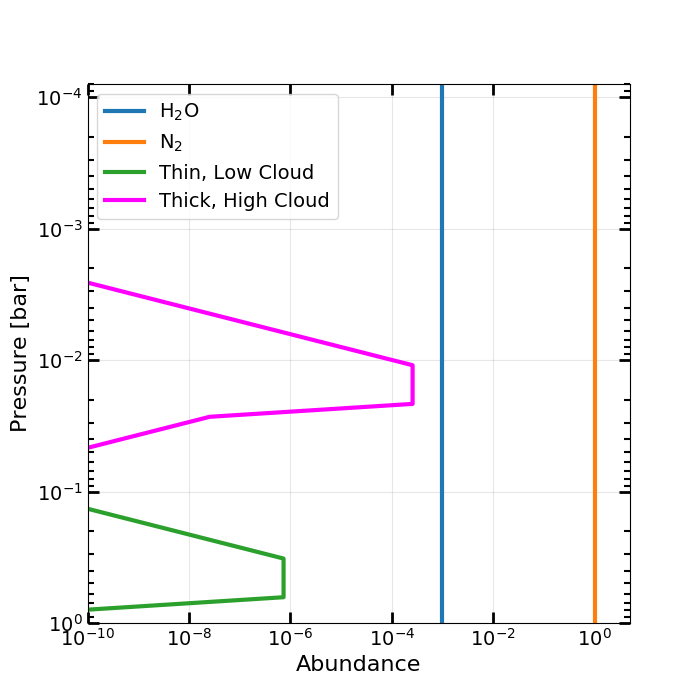}
    \caption{An example of our atmospheric profile, showing a .1\% water vapor atmosphere with no H$_2$ and Earth-like surface pressure. Clouds profiles with top pressures at 10$^{-0.5}$ bars and 10$^{-2.0}$ bars are shown, each with different abundances (column masses of 10$^{-2.5}$ and 10$^{-1.5}$ kg/m$^2$ respectively). The N$_2$ and H$_2$O abundances are volume mixing ratios, while the cloud abundance is measured in mass mixing ratio.}
    \label{fig:atmoprof}
\end{figure}

As a further simplifying choice, we constructed a standard cloud profile to be placed in our atmospheres. This profile, whose vertical location is defined by the pressure level P$_t$ at the top of the cloud, consists primarily of several pressure layers of ``thick'' clouds. Above the cloud top, the cloud particle mass mixing ratio drops rapidly to 10$^{-10}$ over 3.33 scale heights. Beyond this point, no cloud particles are included in the atmosphere. Below the thick cloud ``core'', the mass mixing ratio likewise drops off precipitously to 10$^{-9}$ over 3 scale heights, where it remains down to the planet's surface; see Figure~\ref{fig:atmoprof} for an example atmosphere profile. The mass mixing ratio at any given level is designed such that the sum of cloud mass across all pressure levels is equal to a chosen cloud column mass (hereafter $l$) in units of kg/m$^2$, and yet the cloud is localized at a specific pressure level. Thus, we have reduced our cloud description to two parameters: P$_t$ and $l$.

Through the method described above, we have reduced our rocky exoplanet models from potentially dozens of dimensions to seven. In the process, however, we have discarded some potentially valuable parameters. We plan to gradually restore some of these lost dimensions in future work. Additionally, we recognize that some of the parameters we chose not to explore (such as planet mass) may be determined more easily through other means (such as radial velocity or astrometry measurements).

We expect that the seven parameters (described in Table~\ref{table:params}) adequately describe a rocky Earth-like exoplanet to a first approximation. The combination of these parameters, and the values (shown in Table~\ref{table:params}) chosen for them, gave us a parameter space consisting of 136,000 models. To supplement this, we computed an auxiliary aerosol-free grid, in which all aerosol-related parameters were ignored; thus, noting that we have two ``cloud'' parameters, this auxiliary grid is five-dimensional, containing 5,600 models. The final database thus contains 141,600 model geometric albedo spectra.

\subsection{PSG}
\label{subsec:PSG}

The individual model albedo spectra of our grid were generated using the NASA Planetary Spectrum Generator \citep[PSG;][]{villanueva2018planetary}, an online browser-accessible radiative transfer suite. PSG, developed at the NASA Goddard Space Flight Center, is a comprehensive all-in-one modeling package capable of accurately simulating observations of planetary and sub-planetary-mass objects. It is capable of employing a variety of different molecular line lists including collision-induced absorption cross-sections, as well as pre-computed aerosol scattering models and surface material reflectances, and employs data from sources such as MERRA-2 and MCD to further improve models of Solar System objects. In addition to modeling spectra, PSG also includes software for simulating observations, including observatory throughput and detector noise, and calculates observational uncertainties.

The online, publicly available browser-accessible PSG application is frequently updated with improved functionality. For this project, we used the installable Dockerized version of the PSG application program interface (API); for reference, the installation was performed on 12/13/19. For detailed information about PSG, we recommend \citet{villanueva2018planetary}, or the documentation available on the PSG website (psg.gsfc.nasa.gov). We installed and ran multiple instances of PSG on the Goddard Private Cloud (GPC) at NASA Goddard, and utilized the GPC Job Engine\footnote{https://gitlab.com/mdmoore25404/gpc-je} to distribute calculations across $\sim100$ virtual machines dedicated to the project.

When performing scattering calculations, as required by this work, convergence of the radiative transfer calculations in situations of high optical depth $\tau$ can be challenging. As our model database includes an opacity source that is in some cases extremely dense (clouds), this issue required some special care. For scattering calculations, PSG internally restricts the opacity per layer to be moderate levels ($\tau<1$) in order to ensure convergence of the scattering calculations. Therefore, to avoid this from becoming an issue, we pre-computed the optical depth at each layer within the core of the cloud deck. Any individual layer which we found to have a $\tau \geq1$ was divided into two layers, thus reducing the opacity of each individual layer. Because PSG performs computations including the spherical geometry of the planet's atmosphere, the initial optical depth evaluation was performed near the limb of the planet where the path length of a photon through each layer is greatest.

\section{Results}
\label{sec:database}

\begin{figure*}[hbtp]
    \centering
    \includegraphics[width=7.0in]{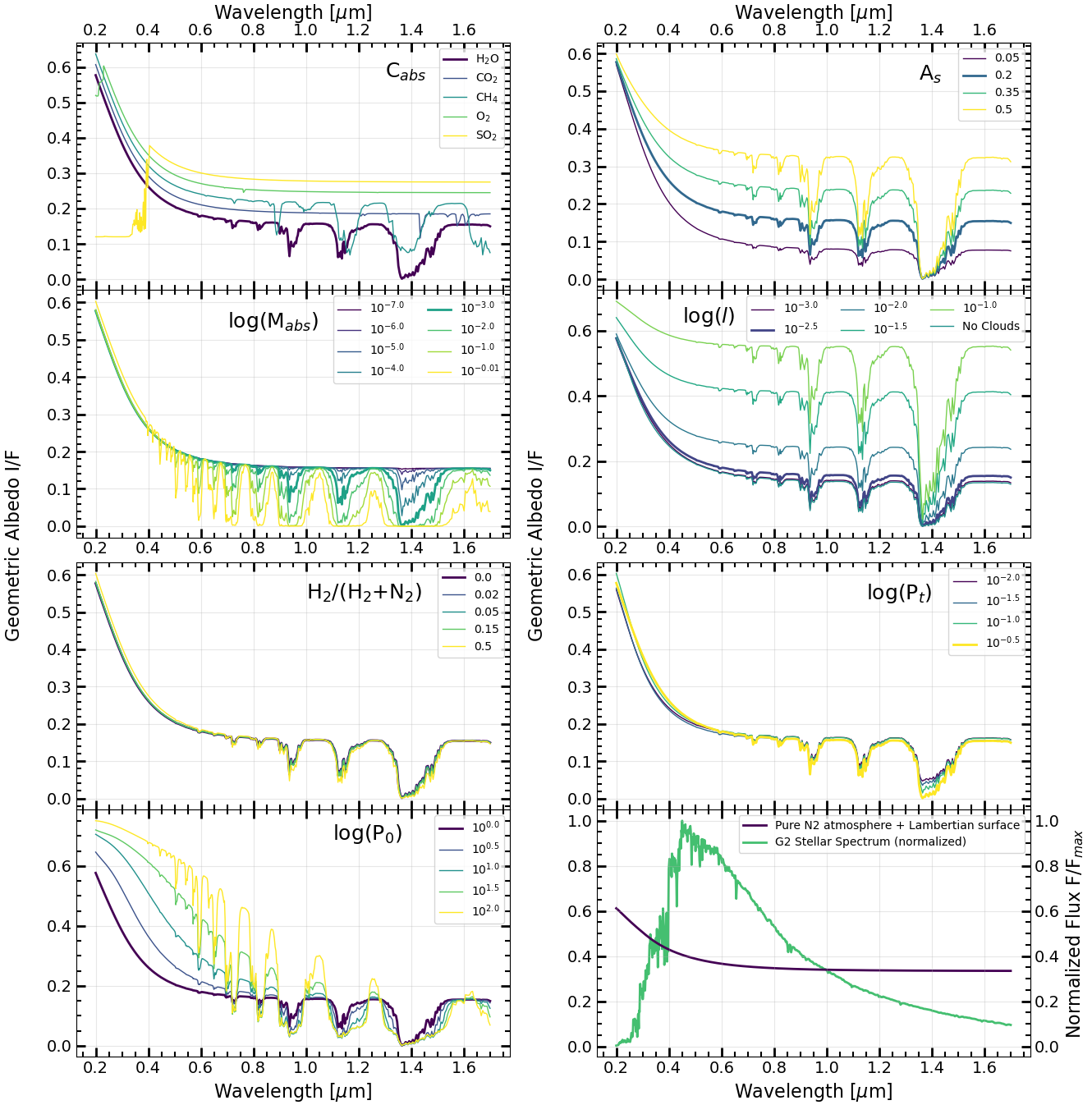}
    \caption{A survey of the variable parameters in our database. The \textbf{bold} line represents the same model, our fiducial model with an H$_2$O/N$_2$ atmosphere, in each sub-panel. The sub-panels show explorations of the dominant absorbing chemical C$_{abs}$ (offset for clarity), its mixing ratio log(M$_{abs})$, H$_2$ content in the background gas, surface pressure log(P$_0$), surface albedo A$_S$, cloud density parameter log($l$), and cloud top altitude P$_t$. A normalized stellar spectrum from a G2 star and the albedo spectrum of a 1-bar N$_2$ atmosphere are included in the bottom right for reference.  See Section~\ref{sec:database} for more details.}
    \label{fig:Database}
\end{figure*}

The full database of geometric albedo spectra for 136,000 cloudy and 5,600 cloud-free model rocky exoplanets is available for download\footnote{ http://doi.org/10.5281/zenodo.3743500}. In addition, the database can used to produce models across the full parameter range of the grid using the REPAST online tool at the Exoplanet Modeling and Analysis Center (EMAC) website\footnote{http://emac.gsfc.nasa.gov/repast}. 

In this Section we will explore some features of the albedo spectra contained within the database. To better highlight the differences between models and to aid in comparisons, we adopt a set of ``fiducial'' parameters. The albedo spectra associated with our fiducial model is present in all figures in this section. This fiducial model has an H$_2$O mixing ratio of $10^{-3}$, and a background gas of pure N$_2$ (such that the parameter H$_2$/(N$_2$+H$_2$) = 0). The surface pressure is set at 1 bar, and the surface albedo is 0.05. We choose a column cloud mass of $10^{-2.5}$ g/kg$^2$, with the cloud top at $10^{-0.5}$ bars.


In Figure~\ref{fig:Database} we plot a sample of the model spectra when varying each of our model parameters.  At the top left, we compare the full spectra for our various absorbing species. It is clear that, at this mixing ratio, CO$_2$ and, beyond 0.3 $\mu$m, O$_2$ have negligible impacts on the spectrum, while many of the H$_2$O and CH$_4$ features in the near infrared region overlap. SO$_2$ has a well-known very strong  absorption feature to the blue of 0.4 $\mu$m. For clarity, these spectra have been plotted to include an artificial offset from each other.


When plotting the absorber mixing ratio for H$_2$O, we see an expected progression from no absorption for extremely low mixing ratios, to very strong absorption for extremely high mixing ratios. Of interest are the increasing number of detectable features in the optical at high mixing ratios, as well as the impact that the changing mean molecular weight has on the effective slope of the spectrum at short wavelengths; this region is dominated by Rayleigh scattering at low absorber concentrations, but drops faster as more short-wavelength flux is absorbed.  Similar distortions of the overall spectral shape are present for the other absorbers.


We find that surface pressure has a significant impact on spectral shapes, as the Rayleigh slope changes dramatically with surface pressure. In addition, high atmosphere pressure has a similar effect on molecular spectral features as increased mixing ratios. The background gas composition has a relatively minor impact at low surface pressure; even when the background gas is 50\% H$_2$, the variation in the spectra is almost undetectable. However, at higher surface pressures than that chosen for our fiducial model, we do note that the effects of H$_2$-H$_2$ collision-induced absorption come into play, creating features beyond 1 $\mu$m.  The three parameters affecting the surface and the aerosol properties (A$_S$, P$_t$ and $l$) all affect the overall continuum shape of the spectra.  Raising the bulk reflectivity of the surface of the planet clearly raises continuum beyond the Rayleigh slope, while having significantly lesser effects on the shape of the Rayleigh slope itself or on the bottom levels of molecular absorption features. The impact of cloud density on the continuum albedo is very similar, but there is a greater effect on the Rayleigh slope and a slightly lesser effect on molecular absorption features in comparison. At high surface pressures and high altitude clouds, the tendency of denser clouds to suppress Rayleigh effects by preventing access to the lower portions of the atmospheres competes with their tendency to raise the overall albedo at all wavelengths.  The impact of cloud top pressure/altitude on the continuum is subtle, with a slight loss of the Rayleigh slope for high-altitude clouds, but the greater impact is on molecular features. High-altitude clouds mask molecular absorption features, as a smaller amount of the atmosphere is accessible. This effect is most noticeable in high-pressure atmospheres.


In comparison to the albedo spectra, we also include a stellar spectrum for a G2 star in Figure~\ref{fig:Database}, normalized to the peak of the stellar output, in order to provide context for the impact of the stellar spectral energy distribution on the planet's observed spectrum (star $*$ albedo). Due to the increased flux at the peak of the stellar spectrum, planetary flux and absorption features in this region will be much easier to detect than features in lower-flux regions. For example, the high absorption of SO$_2$ lies in a region of very low solar output; thus this absorption feature may be difficult to detect due to large observational uncertainties. We also plot a representative albedo spectrum, depicting a 1-bar N$_2$ atmosphere over a 0.5 A$_s$ surface with no clouds or absorbing molecules.

\subsection{Comparison with Solar System Rocky Bodies}
\label{subsec:solarsystem}

\begin{figure*}[hbtp]
    \centering
    \includegraphics[width=5.7in]{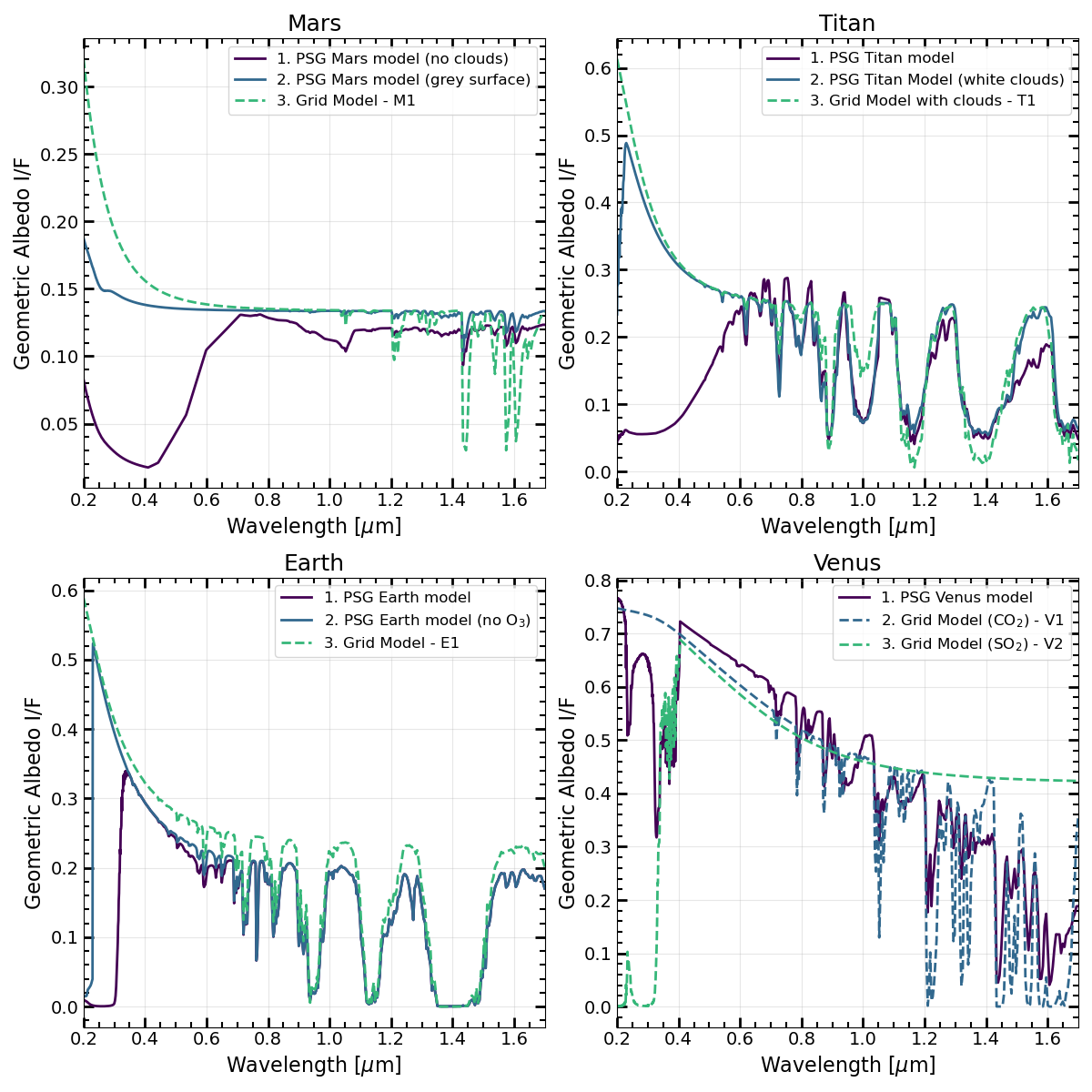}
    \caption{This figure compares each of the atmosphere-rich rocky bodies of the Solar System to one or more analog models from our database for each planet, highlighting the differences between the original, vetted PSG simulation of the planet and our simplified model. For Earth, Mars, and Titan, we also include a transitional model which includes the most significant simplifications but retains much of the complexity of the original model. The parameters for the database analogs are listed in Table~\ref{table:models}.}
    \label{fig:fourplot}
\end{figure*}

\begin{deluxetable*}{cclcccl}[!]
    \tablecaption{Database Analogs to Solar System Planets}
    \tablewidth{0pt}
    \tablehead{ 
    \colhead{Analog} & \colhead{Associated} & \colhead{} & \colhead{} & \colhead{} & \colhead{} & \colhead{} \\
    \colhead{Name} & \colhead{S.S. Planet} & \colhead{C$_{abs}$} & \colhead{M$_{abs}$} & \colhead{P$_0$} & \colhead{A$_s$} & \colhead{Cloud Param.}
    }
    \startdata
    M1 & Mars & CO$_2$ & -0.01 & -1.0 & 0.2 & None \\ 
    E1 & Earth & H$_2$O & -2.0 & 0.0 &  0.35 & None \\
    V1 & Venus & CO$_2$ & -0.01 & 2.0 &  0.2 & log($l$)=-1.5, log(P$_t$)=0.0  \\
    V2 & Venus & SO$_2$ & -5.0 & 2.0 &  0.2 & log($l$)=-1.5, log(P$_t$)=0.0 \\
    T1 & Titan & CH$_4$ & -2.0 & 0.0 &  0.2 & log($l$)=-2.0, log(P$_t$)=-1.0 \\
    \enddata
    \centering
    \tablecomments{All database analogs have no H$_2$.}
    \label{table:models}
\end{deluxetable*}

Since our spectral database examines a range of atmospheric properties for an Earth-mass planet, it is a natural question to ask how these model predictions compare with the albedo spectra of the terrestrial-mass bodies with significant atmospheres in our own Solar System - Venus, Earth, Mars and Titan

However, since our model grid is highly simplified in terms of the atmospheric composition, cloud composition and structure, and surface albedo, any attempt to match the spectra of Solar System planets to the outputs for a specific region of parameter space in the current database is not justified.  In particular, the Solar System terrestrial bodies have a mix of dominant absorbing species: the Venusian atmosphere includes both SO$_2$ and CO$_2$, Titan's spectrum includes absorption from many hydrocarbon species, and Earth's atmosphere obviously includes a number of atmospheric constituents with significant absorption signatures.  Similarly, the cloud properties of Venus and Titan are complex, with combinations of photochemical hazes contributing significantly to the optical spectrum.  Finally, the surface albedo spectrum of Mars has a significant wavelength-dependent reflectivity across the visible spectral region that produces the dominant color contribution to the reflected-light spectrum.  We expressly choose to avoid the added complexities of examining a myriad of specific cloud and haze species and composition-specific surface albedo reflectivities in this study, in order to focus on a small number of variables that are agnostic to the particulars of atmosphere and surface chemistry; we leave an examination of these additional factors for future work, as discussed in Section~\ref{sec:discussion}.

Instead, in order to identify where the atmospheric abundances, surface pressure, cloud properties and average surface albedos for the four atmosphere-rich Solar System rocky bodies would fall within our grid, we have defined analogs within our spectral database, choosing values for these four parameters that most closely approximate the Solar System values. Figure~\ref{fig:fourplot} depicts the spectral differences between the real Solar System bodies and our grid analogs.  It is clear that for Mars and Titan, the dominant change is the drop in flux at short wavelengths; for Mars this is due to the surface reflectivity, while for Titan this is due to the reflectivity of hydrocarbon hazes.  For Earth, our database analog primarily diverges in the near-UV due to absorption from O$_3$; since our grid only has a single absorber in each spectrum, only spectral features of H$_2$O are present.

Finally, for Venus we define two database analogs, since both SO$_2$ and CO$_2$ produce significant absorption across the Venus spectrum. Interestingly, Figure~\ref{fig:fourplot} still shows a difference in the morphology of the SO$_2$ band between the realistic Venus spectrum and our database analog; this is due to the impact of scattering on incoming short-wavelength radiation and the low altitude of SO$_2$ in Venus' atmosphere, which diminishes the absorption.

In Table~\ref{table:models} we list the properties for each of the grid analogs to the Solar System bodies, and the associated label names that we include in subsequent figures. In Section~\ref{sss:analogSS} we will explore what impact of the shift from detailed Solar System planet model to our database analog on our ability to differentiate between them using photometric colors.

\section{Photometric Color Analysis}
\label{sec:observation}

In this Section we use our spectral database to explore the potential to differentiate the properties of rocky planet atmospheres from a multi-band photometric survey. This strategy is motivated by the work of \citet{krissansen2016pale}, hereafter KT16, who showed that producing integrated bandpass filter measurements from optical spectra of Solar System bodies could distinguish potentially Earth-like exoplanets from planets similar to the other (non-habitable) rocky planets in our Solar System. They determined that bandpass ranges of 431-531 nm (``blue''), 569-693 nm (``green''), and 770-894 nm (``red'') could be combined to create sufficient separation in color-color space between Earth and many Solar System bodies.

In Section~\ref{subsec:bandpass} we examine two different sets of photometric bandpasses based on the current expectations for the LUVOIR mission concept, as well as the bandpasses from KT16. We use color-color plots as case studies in Section~\ref{subsec:plots} to consider the efficiency of a purely photometric survey for distinguishing planetary atmospheric characteristics. Section~\ref{subsec:statistics} will review the statistics employed in this analysis.

We note that all of our photometric colors are actually ``albedo colors''- we do not include the impact of the stellar spectrum in the color-color plots.  This is to allow a clear comparison to the example spectra shown in Figure~\ref{fig:Database}. Multiplying a stellar spectrum to the albedo spectra would only apply a multiplication factor to every value, but would not change the statistical analysis of filter positions since they are determined through correlations between colors and various parameters.  The impact of the stellar spectrum is taken into account in Section~\ref{sec:observation} when calculating the actual observable flux for each filter.

\subsection{LUVOIR Photometric Bandpass Case Studies}
\label{subsec:bandpass}



The LUVOIR space observatory concept is one of the major exoplanet direct imaging missions under study for future development; due to the large aperture size, LUVOIR would search hundreds of stars for Earth-like planets as part of an initial reconnaissance survey \citep{luvoir2019luvoir}. We therefore chose to use assumptions about the expected LUVOIR bandpasses for two of our three case studies for a photometric survey. While many features of such a survey are yet to be decided, it is expected that the first step would be the discovery of planets by imaging at a wavelength near the peak photon flux of the planet. As such, we examined an Earth spectrum (generated by PSG using MERRA-2 data) to find this peak flux in a filter bandpass. We found this to be at approximately 0.58 $\mu$m. As the LUVOIR concept would be able to observe in ultraviolet, optical, and a red or near-infrared channel simultaneously, we paired this with a second photometric point at 0.4 $\mu$m. We justify this wavelength choice in Section~\ref{subsec:statistics}. Finally, we paired these two with a third filter choice at an indeterminate point red of 0.58 $\mu$m. For Case 1, we assumed the photometric bandpasses of the filters would be 10\%, based on the expected achievable bandpass width for an on-axis telescope design baselined for the 15-meter LUVOIR-A concept \citep{luvoir2019luvoir}.

The second case study is identical to the first, except that the width of the filter bandpass has been changed to 20\%. This bandpass would be achievable with an off-axis design, as baselined for the smaller LUVOIR-B concept.



\subsection{Statistical Methodology}
\label{subsec:statistics}

For the LUVOIR bandpasses, we employed the Pearson correlation coefficient and Grubbs's test for outliers to guide our decisions of bandpasses to explore in color-color space. 

The Pearson correlation coefficient r is a measure of how close the data points lie to a linear correlation; $-1\leq$ r $\leq 1$ where r $=\pm 1$ indicates a perfectly (positive or negative) linear correlation and r$=0$ indicates no correlation. The Pearson correlation coefficient r$_{xy}$ between variables x and y can be calculated for a sample by

\begin{equation}
    r_{xy} = \frac{\sum_{i=1}^{n}{(x_i-\bar{x})(y_i-\bar{y})}}{\sqrt{\sum_{i=1}^{n}{(x_i-\bar{x})^2}}\sqrt{\sum_{i=1}^{n}(y_i-\bar{y})^2}}
\end{equation}

where n is the sample size, x$_i$, y$_i$ are individual sample points, and $\bar x$, $\bar y$ are the sample means for x and y respectively. Interpretation of the Pearson coefficients requires some caution as they do not capture such information as the `slope' of the correlation. However, Pearson correlation coefficients are useful for gaining some intuition for the value of each filter. 

Grubbs's test for outliers, as the name suggests, is useful for locating unusual data points in a sample. The one-sided Grubbs test statistic G is defined as

\begin{equation}
    G=\frac{y_{max}-\bar y}{s}
\end{equation}
\nopagebreak
where s is the sample standard deviation, y$_{max}$ is the largest value in the sample, and $\bar y$ is defined as above. Then, if 

\begin{equation}
    G>\frac{N-1}{\sqrt{N}} \sqrt{\frac{t_{\alpha/N,N-2}^2}{N-2+t_{\alpha/N,N-2}^2}}
\end{equation}

where N is the sample size and $t_{\alpha/N,N-2}^2$ is the upper critical value of a t-distribution, we reject the hypothesis that y$_{max}$ is not an outlier with a significance level $\alpha$. We then remove $y_{max}$ from the sample and repeat the test until no outliers are found.

To determine the best filter positions to distinguish the impact of each specific parameter, we calculated the Pearson correlation coefficients for the color of different filters combined with our 0.58 $\mu$m central filter, and then applied Grubb's test in order to help find ``abnormally strong'' correlations. Like the Pearson correlation coefficient, the Grubbs test should only be regarded as a guide. It was used only to highlight potentially useful combinations for further study.

\subsection{Survey Results}
\label{subsec:plots}

In the following section, we will review the results of the three cases laid out in Section~\ref{subsec:bandpass}. In each of the figures below, we highlight five models in our grid chosen to approximate the atmosphere-rich terrestrial bodies within our Solar System. Details of these five models are given in Table~\ref{table:models} and Section~\ref{sss:analogSS}.

\begin{figure}[ht!]
    \includegraphics[width=3.3in]{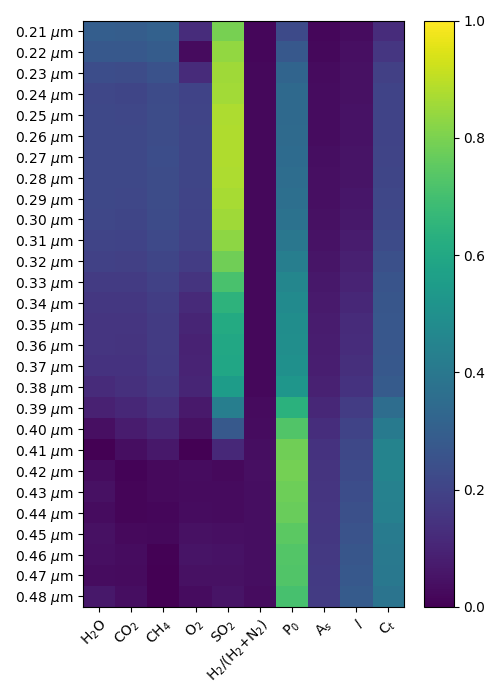}
    \caption{The absolute values of Pearson correlation coefficients calculated for individual 10\% filters in the UV band with a variety of central wavelengths. These measure how well a change in the parameter value correlates with a change in the filter response. Values close to 0 indicate very little or no correlation, while values close to $\pm 1$ indicate very strong, almost linear correlations. Notice the shift in correlation emphasis from SO$_2$ concentration to atmospheric and cloud parameters around 0.4 $\mu$m. We chose this location for the center of our UV bandpass as mentioned in Section~\ref{subsec:bandpass}.}
    \label{fig:PearsonUV}
\end{figure}

\begin{figure}[t!]
    \includegraphics[width=3.3in]{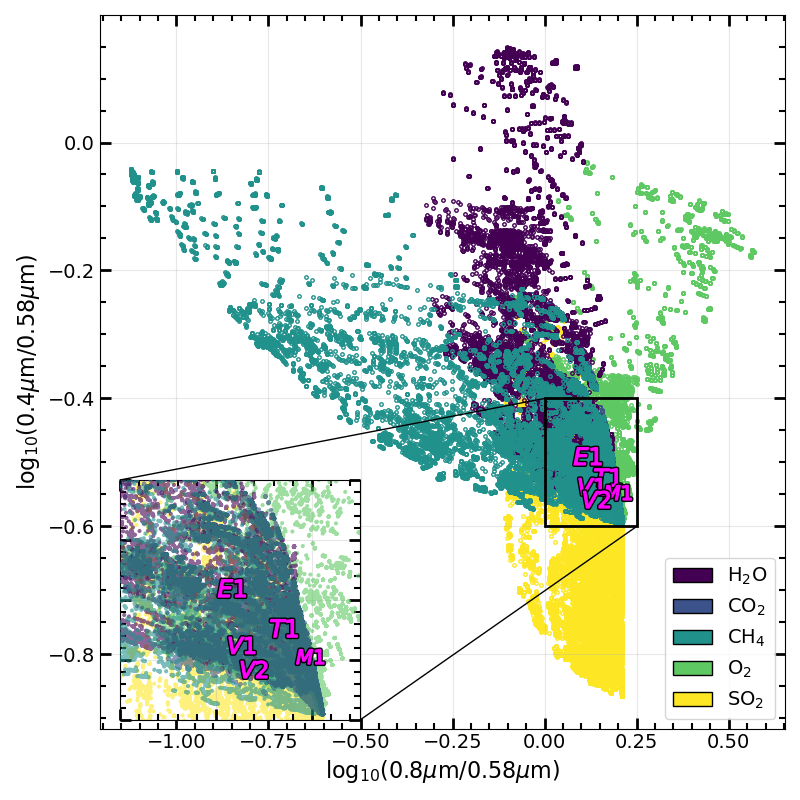}
    \caption{Our model database, and several Solar System planets, plotted on our chosen axes. Each axis is the logarithm of the ratio of two 10\% filters, centered at the indicated wavelengths. Here we have highlighted the dominant absorbing chemical in each model on the plot. We note the four distinct branches separating SO$_2$, O$_2$, H$_2$O, and CH$_4$. On these axes, CO$_2$-dominated atmospheres fail to manifest a distinct branch due to a lack of spectral features in any of the bands.}
    \label{fig:case1}
\end{figure}

\subsubsection{LUVOIR Bandpasses: Filter Position Selection}
\label{sss:LUVOIRfilterchoices}


For our UV filter options, the only parameters that showed significant Pearson correlation coefficients were SO$_2$ (due to the strong absorption feature) and surface pressure (due to Rayleigh scattering); see Figure~\ref{fig:PearsonUV}) for a visual representation of the Pearson coefficients. In order to be sensitive to the impact of both parameters, we decided to choose a filter position at 0.4 $\mu$m for both our 10\% and 20\% cases, since it covers a region of reasonable sensitivity to both parameters.

For the 10\% filter case, we found that a third filter located at 0.80 $\mu$m, in combination with the 0.4 and 0.58 $\mu$m filters, produced an outlier-level correlation for four of the five chosen molecular species concentrations. We note, however, that this correlation was still quite low (approximately 0.2-0.25). For our 20\% filter scenario we found that a similarly located filter to that chosen in the 10\% scenario, at 0.78$\mu$m, yielded similar - though diluted - results.

In addition to finding outliers in our 0.8 and 0.78 $\mu$m filters, which we ultimately chose as our final selections, we also noted several other filters which correlated well with several parameters. For example, blue and green optical filters (at 0.58, 0.66, and 0.68 $\mu$m) correlated strongly with cloud density when combined with our pre-chosen filters, while some filters in the near-infrared showed correlation with O$_2$ as well as CH$_4$ (at 1.34 and 1.36 $\mu$m) and H$_2$O (at several locations between 1.38 and 1.46 $\mu$m). However, we did not choose these as they were only able to distinguish two or, in some cases, three of the five molecules. We did not find any strong or outlier correlations for CO$_2$ in our study. In each case, we plotted the filter combinations in color space in order to see how distinct (or degenerate) the correlations between different parameters was. Our final choice of the 0.8 and 0.78 $\mu$m filters for 10\% and 20\% respectively was made based on achieving the maximum color-color correlation for the most parameters.

\subsubsection{LUVOIR Bandpasses: Color-Color Plots}
\label{sss:cases1n2}

\begin{figure}[!]
    \includegraphics[width=3.3in]{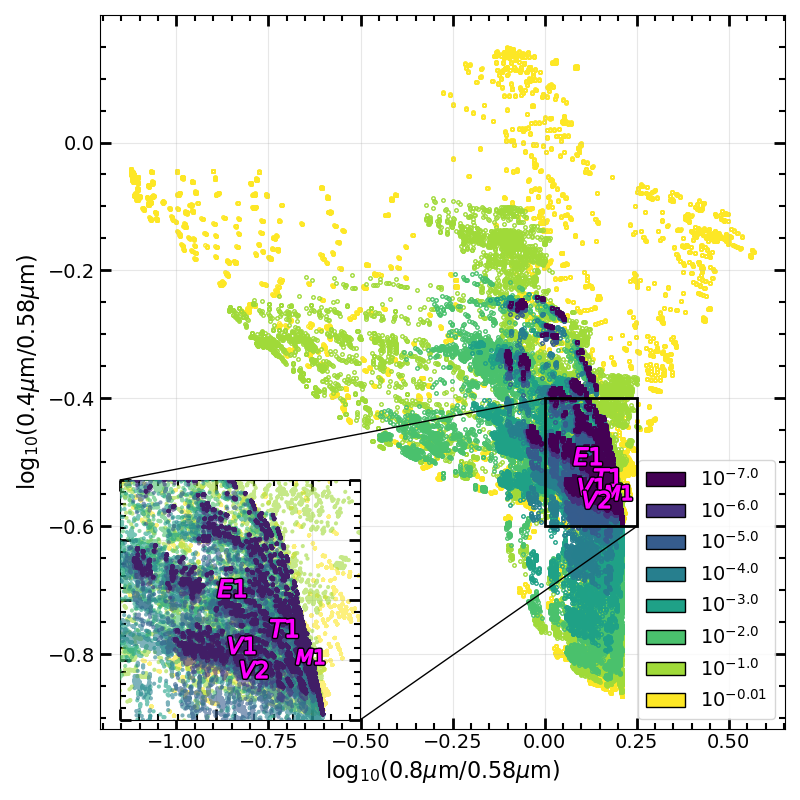}
    \caption{As Figure~\ref{fig:case1}, however in this plot we have highlighted the mixing ratio of the primary absorber. Notice that the H$_2$O and O$_2$ mixing ratio must exceed 0.1, and CH$_4$ mixing ratio must exceed 0.01, before the models move away from the central grouping of models. This suggests that photometric characterization will only be effective for identifying atmospheres with a high concentration of molecular absorbers.}
    \label{fig:mixrats}
\end{figure}

\begin{figure}[t!]
    \includegraphics[width=3.3in]{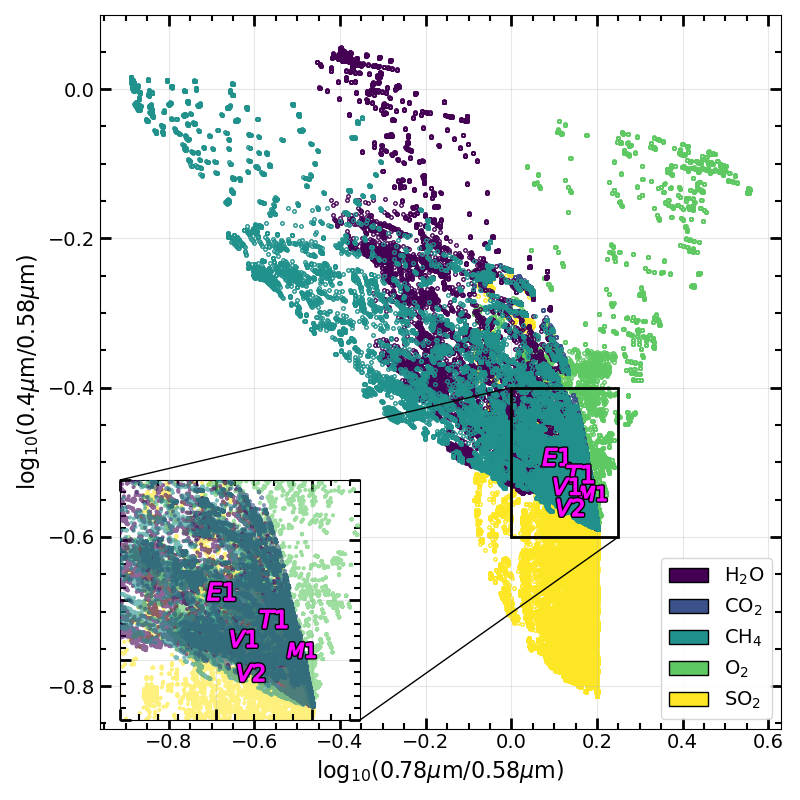}
    \caption{As Figure~\ref{fig:case1}, however in this plot we use 20\% bandpass filters. The results are very similar to Figure~\ref{fig:case1}. However, note that H$_2$O and CH$_4$ are more mixed and harder to distinguish in this case. The difficulties observed in Figure~\ref{fig:mixrats} are also present here.}
    \label{fig:case2}
\end{figure}

\begin{figure}
    \includegraphics[width=3.3in]{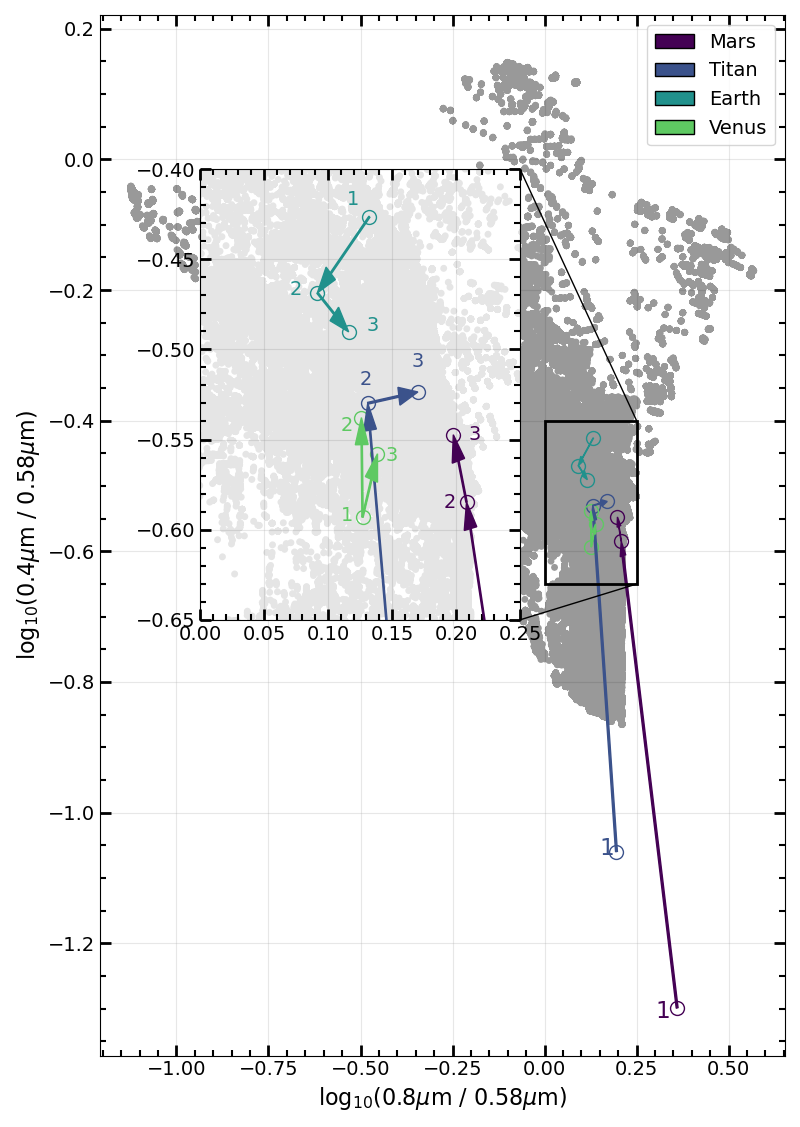}
    \caption{Here, we highlight the shift in color-color space from each of the terrestrial solar system planets to the analogs referenced in the paper. As shown, the differences are extremely minor for the relatively simple Earth and Venus models. Titan and Mars shift in a more significant way, as shown in Figure~\ref{fig:fourplot} and discussed in Section~\ref{sec:discussion}. The axes here are the same as those used in the 10\% bandpass case. The numbers adjacent to each point refer to the numbered models in Figure~\ref{fig:fourplot}. }
    \label{fig:transit}
\end{figure}

Figure~\ref{fig:case1} shows a color-color plot of our grid using our three chosen 10\% filters, as well as our Solar System database analogs. Note that each of H$_2$O, CH$_4$, CO$_2$, and SO$_2$ appear to follow their own branch on this plot. However, the separation from the central ``group'' of data points does not manifest until a significant concentration of the chemical is present in the atmosphere, as indicated in Figure~\ref{fig:mixrats}. In addition, we observe that our Solar System analog objects all lie in the central region where the branches overlap; the only analog S.S. planet with any appreciable separation is E1, which is due to the absorption from H$_2$O.

Figure~\ref{fig:case2} shows a color-color plot for our chosen 20\% filters. This set shows similar trends as those observed in Figure~\ref{fig:case1}; most notably, there are four branches corresponding to H$_2$O, CH$_4$, CO$_2$, and SO$_2$. However, there is a greater degree of degeneracy between H$_2$O and CH$_4$ on these axes, and the same clustering of solar system analog objects.

\subsubsection{Exploration of Solar System Analogs in Color-Color Space for 10\% Bandpasses}
\label{sss:analogSS}

In Figures \ref{fig:case1}, \ref{fig:mixrats}, and \ref{fig:case2}, it is clear that our Solar System database analogs are tightly clustered and difficult to differentiate relative to other models covering a variety of atmospheric properties. It is instructive to explore this phenomenon and how it compares against the positions of real (non-analog) Solar System objects on these axes.

Figure~\ref{fig:transit} traces the movement in color-color space for each Solar System planet as the model parameters are adjusted to match the chosen analog models from our database; the numbers proceed from the realistic models of Earth, Mars, Titan, and Venus to our E1, M1, T1, and V1 and V2 analog objects, and are matched to the numbers of the spectra presented in Figure~\ref{fig:fourplot}. It is clear that Earth and Venus have their positions shifted only slightly by the transition, since both planets' spectra are dominated by clouds that are featureless across the visible region of the spectrum, and the key wavelength-dependent variation is driven by atmospheric absorption. 

In contrast, the positions of Mars and Titan are changed significantly compared with our database analogs.  For Mars, this is due to the removal of the detailed surface reflectivity present in the realistic Mars spectral model, while the presence of tholin hazes significantly affect the spectral color of Titan.  As stated earlier, we chose not to include either of these factors in the current database to preserve simplicity and a minimal parameter set; we leave this work to future studies. 

\subsubsection{Bandpasses from Krissansen-Totton et al.}
\label{sss:case3}

\begin{figure}[t!]
    \includegraphics[width=3.3in]{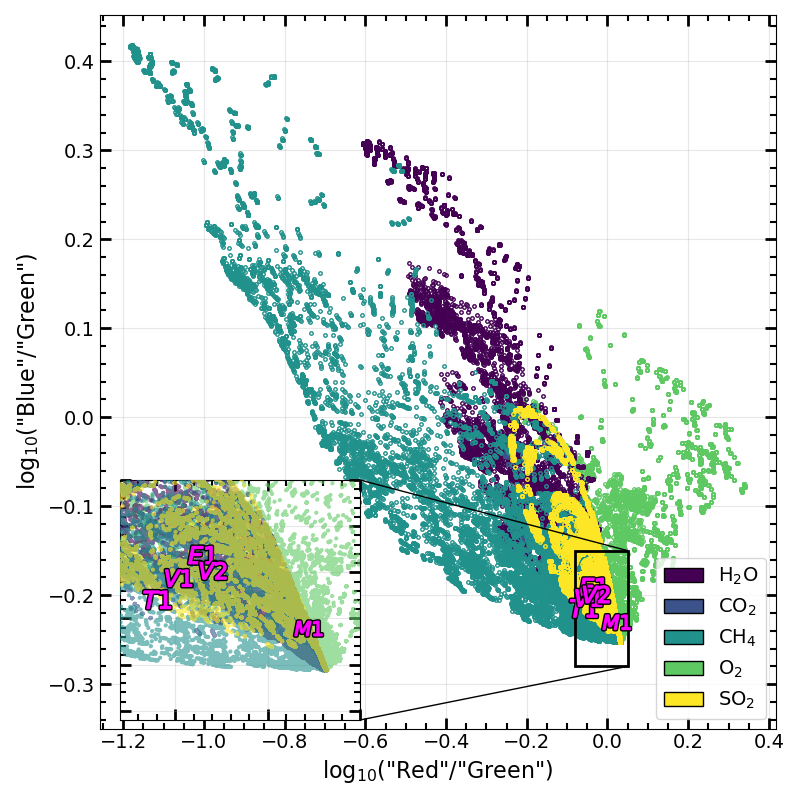}
    \caption{Our database is plotted on the same axes as used in Figure 2 in from \citet{krissansen2016pale}. Note the strong clustering of our Solar System analog models on these axes, as well as the lack of a distinct SO$_2$ branch.}
    \label{fig:KT16}
\end{figure}

Figure~\ref{fig:KT16} shows our database plotted on the red-blue-green axes chosen by KT16 as described in Section~\ref{subsec:solarsystem}. It is notable that our Solar System analogs models are very tightly clustered on these axes, even more-so than in Figure~\ref{fig:case1} and Figure~\ref{fig:case2}, except now the Mars analog is slightly separated.  We also note that the KT16 bandpasses reduce the overlap between the methane and water ``branches'' at lower concentrations. However, the SO$_2$ branch is completely eliminated, and the O$_2$ branch is shorter in both axes.

\subsubsection{Estimating Observing Time for Discriminating Atmospheric Properties of Rocky Exoplanets}
\label{sec:obstime}

\begin{figure*}[hbtp]
    \centering
    \includegraphics[width=6.0in]{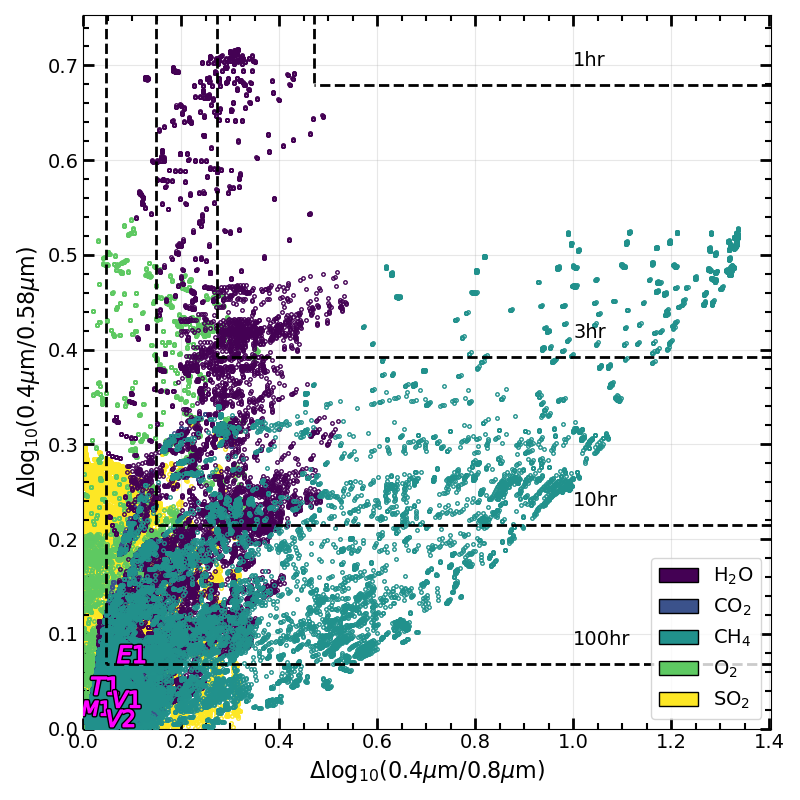}
    \caption{In this plot, a G2 stellar spectrum has been applied to the raw albedo spectra prior to calculating spectroscopic data from 10\% filters, in order to approximate actual planetary fluxes for a planet around a solar-type star. Further, the X- and Y-axes have been changed to show the absolute difference of the filter response ratio from that of a cloud-free, pure N2 atmosphere at 1 bar P$_0$. Included are the 3-$\sigma$ noise floors for a planet at 10 parsec for 1, 3, 10, and 100 hour observations with a 15m telescope, calculated with PSG (assuming zero detector noise), which indicate how difficult it would be to distinguish between a given model and a simple world with no absorbers. Letters indicate the position of the Solar System planet analogs described in Section~\ref{subsec:solarsystem}.}
    \label{fig:solarplot}
\end{figure*}

In order to assess the effectiveness with which different planetary atmosphere compositions can be distinguished for a specific observatory and instrument suite, it is necessary to actually calculate the measurement uncertainty of observations in the chosen photometric bands, and compare that to the expected difference in photometric color for different models with different atmospheric compositions.

In Figure~\ref{fig:solarplot}, we have multiplied a G2 stellar spectrum with our raw model albedo spectra to determine real photometric colors in the LUVOIR 10\% bandpasses, and then calculated the difference in color of that model compared with a cloud-free, pure N2 atmosphere with 1-bar surface pressure; this allows us to evaluate how well a certain model could be distinguished from an absorber-free atmosphere.  We also calculated the photon-limited uncertainty in the determination of each color for an observation of a solar-type star at 10 parsec with a 15-meter telescope at a variety of integration times using PSG and the telescope and instrument parameters provided in the LUVOIR study report \citep{luvoir2019luvoir}. We chose to use a ``perfect'' detector for this calculation, in order to maintain applicability to multiple future observing projects; thus, no detector noise, read noise, dark current, or other sources of uncertainty are considered here. As such, the noise floors depicted are likely optimistic.

Under these conditions, we found that a 3-hour minimum observation would be required to begin distinguishing the H$_2$O and CH$_4$ models from a pure-N$_2$ atmosphere. With a 10-hour observation, atmospheres with high fractions of O$_2$ and SO$_2$ could be identified.  Finally, with a 100-hour observation, a planet with Earth's H$_2$O abundance could be positively identified.



\section{Discussion}
\label{sec:discussion}

We have computed a database of 141,600 geometric albedo spectra for rocky Earth-mass exoplanets. Individual planet spectra can be derived from the database using the online calculator hosted at the EMAC website\footnote{http://emac.gsfc.nasa.gov/repast}, and the whole database is publicly available\footnote{http://doi.org/10.5281/zenodo.3743500}.

Using this database, we simulated a photometric color-color analysis using a variety of filter combinations. We examined photometric filters with 10\% and 20\% bandpasses in the ultraviolet, visible, and near-infrared. We found that, given a plausible discovery bandpass at 0.58 $\mu$m and a secondary ultraviolet bandpass at 0.4 $\mu$m, adding a third filter bandpass at 0.8 $\mu$m (for 10\% bandpasses) or 0.78 $\mu$m (for 20\% bandpasses) allowed discrimination between four of the five molecular absorbers simulated in this study. We note however that exceptional concentrations of the chemical species we examined are required to distinguish planets using shorter integration times; in addition, we did not include all sources of instrument noise, and therefore the results could be even less promising than described here.

As mentioned in Section~\ref{subsec:design}, we made a number of other simplifying assumptions in the interest of computational feasibility. We intend to address several of these assumptions in future expansions on this database, beginning with the choice of a gray surface albedo. While this simplifying assumption is common in the literature and is reasonable as a first approximation, several studies \citep{madden2020surfaces, o2018vegetation, schwieterman2015nonphotosynthetic, kaltenegger2007spectral} have shown that the surface composition of rocky exoplanets can have a significant effect on reflected light spectra. The impact of using a grey surface albedo is also quite notable in comparison to objects in our Solar System; see for example Figures \ref{fig:fourplot} and \ref{fig:transit}. However, the addition of appropriate surfaces would correct this discrepancy, and allow a model database to more appropriately cover the regimes that Solar System objects inhabit. Because of this, and other comparisons (e.g. to Earth), expanding our database to cover a variety of surfaces is a very high priority for the further development of our models.

Our list of dominant absorbers did not include molecular species that are produced primarily by photochemistry in cool rocky planet atmospheres. In particular, O$_3$ is a byproduct of O$_2$ photochemistry, and is responsible for a significant spectral absorption feature in Earth's spectrum.  Similarly, we did not include CO, or various chemical and photochemical byproducts of CH$_4$, which are present in the atmospheres of Venus and Titan. We decided not to include these spectral absorbers since their abundance is highly dependent on a planet's bulk atmospheric composition and the spectral energy distribution of a planet's host star, and therefore the parameter space where these absorbers would be important will be restricted. For this study, we were focused on bulk constituents that are expected to be present in the majority of cool rocky planet atmospheres. We intend to explore ways to include more realistic chemistry into future spectral databases.

We included molecular hydrogen H$_2$ as a potential background gas component in this database, since it has been suggested \citep[][e.g. in]{ramirez2014warming} that Mars retained an atmosphere with significant amounts of H2 briefly after formation, with climate implications \citep[see also][for implications for close-orbiting planets.]{koll2019hot}. However, the detection of H$_2$ in terrestrial planet atmospheres primarily relies on H$_2$-H$_2$ collision induced absorption, and our results show that H$_2$-H$_2$ CIA absorption does not become apparent until concentrations and/or pressures become fairly extreme. Thus, this parameter may be restricted in future database releases in favor of other, more impactful dimensions.



Additional areas of future work include variations in aerosols; our study was limited to a featureless cloud by computational necessity. However, the addition of high-altitude photochemical hazes are also worthy directions to explore, as these are common in rocky planet atmospheres within the Solar System and have an impact comparable to that of planetary surfaces. An exploration of other particulate matter (such as dust) suspended in the atmosphere is also under consideration. Also, potential follow-ups to the photometric analysis presented in this paper include the application of machine learning tools \citep[as in][]{batalha2018color} to make the filter choices more agnostic.

\acknowledgements
    The authors would like to acknowledge Y. K. Feng and Dr. Joshua Krissansen-Totton for providing albedo spectra for validation; Dr. Natasha Batalha for guiding us towards useful analysis techniques; and Dr. Chris Stark and Dr. Matthew Bolcar for descriptions of expected LUVOIR spectroscopic approaches. The material is based upon work supported by NASA under award number 80GSFC17M0002.

\software
    SciPy \citep{virtanen2019scipy}, NumPy \citep{van2011numpy}, matplotlib \citep{hunter2007matplotlib}, pandas \citep{mckinney2011pandas}, astropy \citep{astropy:2013,astropy:2018}, PSG \citep{villanueva2018planetary}

\bibliography{main.bbl}

\end{document}